\documentclass[12pt]{article}\usepackage[dvips]{graphicx,color}
\def\bea{\begin{eqnarray}}
\def\eea{\end{eqnarray}}
\def\be{\begin{equation}}
\def\ee{\end{equation}}
\def\nn{\nonumber}

\def\G{\Gamma}
\begin{document}
\begin{center}
{\large\bf Semiclassical Horizons}\\
\vspace{2cm}

{\sc Arundhati Dasgupta \footnote{E-mail: adasgupt@unb.ca}}\\
{Department of Mathematics and Statistics,\\}
{University of New Brunswick,\\}
{Fredericton, E3B 5A3, Canada.}

\end{center}
\vspace{2cm}

\noindent
{\bf Abstract:}
The entropy of apparent horizons is derived using coherent states or semiclassical states in quantum gravity.
The leading term is proportional to area
for large horizons, and the correction terms differ according to the 
details of the graph which is used to regularise the quantum gravity phase space variables.

\newpage
\section{Introduction} 

A black hole space-time is characterised by parameters like mass, angular momentum and charge. This behaviour is similar to
thermodynamic systems like that of a ideal gas, where details of the dynamics of the molecules is averaged over
and the entire system is described by macroscopic parameters like temperature, volume, pressure etc.
Further, the laws of black hole mechanics are very similar to laws of thermodynamics,
including a second law which says the area of horizon increases in any physical process.
This led to the conjecture that horizons have entropy like a thermodynamic system, and
the entropy is equal to area of the horizon divided by four Planck length squared
($A_H/4 l_p^2$). This conjecture is not proven experimentally though for theoretical
consistency, a quantum theory of gravity is searching for a microscopic origin of horizon entropy.

\noindent
Black holes like the
Schwarzschild space-time are vacuum solution of Einstein's equation. One
cannot say that these are comprised of fundamental constituents like the gas molecules
and the averaging over these degrees of freedom give the macroscopic entropy
and temperature. Even if the fundamental structure of space-time is quantised, then it has to be explained why the
black hole is not a pure `condensate' of such fundamental degrees of freedom. Thus it appears that a plausible explanation is the entropy
has originated from tracing over a part of the system. There clearly is a loss of information for `a part of the system' for the
classical outside observer, due to the presence of the horizon. In a quantum mechanical
description of the black hole space-time, if the outside and the inside of the
horizon are described by two different Hilbert spaces, and the outside observer,
has access to one of them, one has to trace over the Hilbert space
inside the horizon. A density matrix then describes the system outside the horizon. 
If the two Hilbert spaces inside and outside the horizon are 
entangled, then the reduced density matrix is a mixed one and a entropy results. Thus the task would be to find a 
`quantum mechanical' wavefunction for the horizon and trace over the
part of the wavefunction within the horizon. 
Since horizons are macroscopic and lightcones which determine the causal
properties of any space-time and are classically well defined, any attempt
to obtain a `exact quantum black hole' would require non-perturbative
`solitonic' sectors to exist in the quantum theory, or they should be
recovered from semi-classical states just as light waves emerge from
coherent states which are condensates of photons described by the electromagnetic `quantum'
states. 

\noindent
Loop quantum gravity (LQG) is a approach to quantum gravity, where the canonical
metric variables and the extrinsic curvature is redefined as a SU(2) gauge connection,
and it's conjugate momentum or electric field. A kinematic Hilbert space can be identified
for this theory, though the constraints have to be solved yet. A coherent state
can be described in the kinematic Hilbert space, which have all the properties of the
usual coherent states including minimum uncertainty, overcompleteness, and peakedness
about classical solutions \cite{hall,thiem1,thiem}. While this is not a 
complete description, as the coherent states are defined in the kinematic
Hilbert space, certain questions about quantum fluctuations about the classical
geometry, resolution of the singularity at the center of the black hole, and correlations across the horizon can be answered.

\noindent
The LQG Hilbert space uses regularised canonical variables, defined
along one dimensional piecewise analytic edges (e)  of a graph ($\Gamma$).
The configuration space variables are the holonomy $h_e$ of a SU(2) gauge 
connection comprising of the tangent space `spin connection' and the extrinsic curvature
of the spatial slice. The `dual' momentum $P^I_e$ (I=1,2,3 is the SU(2) index) is the set of triads describing
the intrinsic geometry of the spatial slice, smeared in two dimensional 
surfaces, which the edges intersect. The pair $h_e, P^I_e$ comprise the
phase space variables, and one can define a coherent state for each
edge in the graph $\psi_t(g_e h_e^{-1})$, peaked at the classical values of the complexified version of these variables $g_e=\exp(i T^I P^I_{e})h_e$,
where $T^I$ are the generators of SU(2)
\cite{thiem1}. The semiclassicality parameter $t$ controls the quantum fluctuations. This parameter is fixed as $t=\frac{l_p^2}{A_H}$ where $l_p$ is the Planck length, and
$A_H$ is the area of the horizon associated with a black hole. Thus for large black holes $t$ is small, and the semi-classical approximation is better, and for
Planck size black holes, the semiclassicality parameter $t$ is order 1, and hence one has to fully quantise the system in that regime.  Once a graph is obtained discretising the entire spatial slice, as observed in \cite{adg1,adg2,adg3}, the regularised
variables are well defined even at the central singularity of the
black hole, and smooth across the horizon for a particular spatial
slicing.

\noindent

\section{Entropy of Horizons}
The coherent state is defined for a graph which when embedded in the classical space-time comprises of edges along the
coordinate lines of spherically symmetric axis ($r,\theta,\phi$). 
The apparent horizon equation $ \nabla_a S^a - K_{ab}S^a S^b -K=0$ ( $S^a$ is the normal to the apparent horizon,$K_{ab}$ the extrinsic curvature
of the spatial surface and K, the trace of the extrinsic curvature) is re-written in the
regularised `classical variables' is satisfied in the classical limit,
and encodes correlation across the horizon \cite{adg2}. Written in terms of the holonomies at a vertex $v_1$
outside the horizon, and $v_2$ inside the horizon, one obtaines

\begin{eqnarray}
&&4P_{e_\theta}^2 \left[{\rm Tr}\left(T^J h_{e_{\theta}}^{-1} V^{1/2} h_{e_{\theta}}\right)_{v_1}- {\rm Tr}\left(T^J h_{e_\theta}^{-1}V^{1/2}h_{e_\theta}\right)_{v_2}\right]{\rm Tr}\left(T^J h_{e_\theta}^{-1}V^{1/2}h_{e_\theta}\right)_{v_1} \nn \\
&&-\frac{1}{\sqrt \beta} \frac{\partial}{\partial \beta}{\rm Tr} \left(T^I \ ^{\beta} h_{e_\theta}\right) P^{I}_{e_\theta}\ _{v_1} =0 
\label{diff}
\end{eqnarray}
$e_{\theta}$ denotes a edge along the coordinate lines of $\theta$, $\beta$ denotes the Immirzi parameter, $h_{e_\theta}$ and $P^I_{e_\theta}$ denote the holonomy
and momenta along the edge $e_{\theta}$, one set for a edge beginning/ending at vetex $v_1$ another set for a edge beginning/ending at vertex $v_2$.
Symbolically it has the following form
\be
\hat A[\hat B^I(v_1)- \hat B^I(v_2)]\hat C^I(v_1)- \hat D =  0 
\ee
where $\hat A,\hat B,\hat C,\hat D$ are operators given as functions of $h_e,P^I_e$
of the angular edges which begin/end at a vertex $v_1$ outside the horizon and those
angular edges which begin/end at a vertex inside the horizon. In the coherent states
this can be written as\be
{\rm Limit}_{t\rightarrow 0}<\Psi|\hat H|\Psi>=0,
\ee
The expectation value of the horizon operator is thus vanishing only in the semi-classical limit, realised as
$t\rightarrow0$.
where $ 
\hat H=\hat A \hat B^I (v_1) \hat C^I(v_1) - \hat A\hat B^I(v_2)\hat C^I(v_1) -\hat D.$

{\it Note that this set of equations show that the coherent state wavefunction within the horizon must be correlated with the 
coherent state outside the horizon. The correlation was encoded in a conditional probability function to derive what the
zeroeth order entropy should be, but the actual computation of the correlations and the density function should be obtained
using physical coherent states, a first step towards which has been taken in a recent paper by Thiemann et al \cite{thiem}}

\includegraphics[scale=0.5]{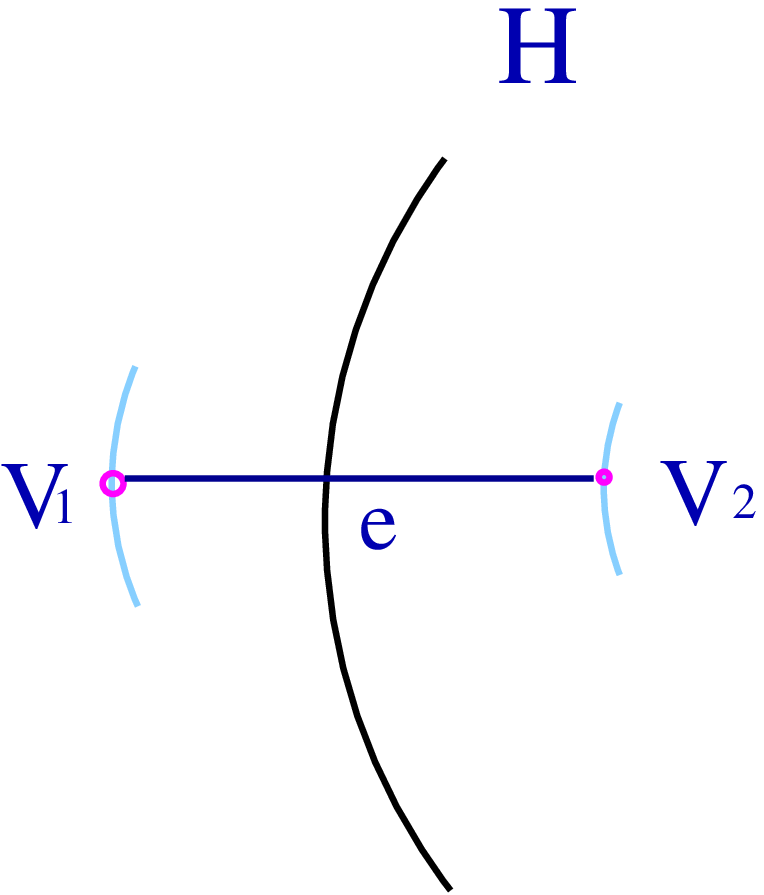}

In the above, the graph at the horizon is taken to simply comprise of
radial edges linking vertices inside the horizon to those
outside the horizon. 
Further the coherent states of the angular edges (along the $\theta,\phi$ coordinates) ending/beginning at vertices $v_1$
outside the horizon get correlated with coherent states of angular edges ending/beginning
at vertices $v_2$ inside the horizon. Note $v_1$ and $v_2$ are linked
by a radial edge $e_H$. Thus the coherent state is written thus
\be
|\Psi> = \prod f(v_1, v_2) |\psi_{v_{1}}>|\psi_{H}>|\psi_{v_2}>
\ee
A tracing over of the edges inside the horizon (or the set of coherent states in the above tensor product state labelled by $v_2$),
gives a mixed density matrix $\rho$, which in the classical limit is diagonal \cite{adg3}.
The entropy obtained as $-\rm Tr(\rho \ln \rho)$ of this density matrix,
is the number of ways to induce the horizon area as the degeneracy due to the horizon state
$|\psi_{H}>$ contributes to the trace (The degeneracy of a horizon state which induces a area $(j_e+1/2)t$ is $2j_e+1$. Thus given a set of radial edges inducing the horizon with area, one simply
obtaines the degeneracy associated with the coherent state for those edges, given the total area of the horizon, and 
a log of that gives the entropy.

{\it Note that this way of deriving entropy of log of the spins at the horizon has been used, and is similar to the `it for bit'
formulation introduced by Bekenstein, however, this is a quantum mechanical derivation of the similar principle using a coherent state
wavefunction. Also this differs from the derivations of \cite{log} as there is no additional constraint to restrict the degeneracy of area at the horizon. In my formulation,
the horizon wavefunction is free as per the apparent horizon equation (\ref{diff}) in the semiclassical limit.}

\noindent
Thus the degeneracy of the horizon coherent state at the horizon is determined by the number of edges, as well as the
spins associated with those edges.  
This brings us to the {\it choice of the graph at the horizon}, which
will determine the number of ways to induce the horizon area. Having
only one graph, with the edges each carrying spin $j_e$, with the area equally distributed 
on the horizon sphere one can count the entropy as simply by summing the degeneracy $(2j_e+1)$
associated with each edge. The constraint is that if N is the number of edges, $(j_e+1/2) N=\frac{A_{H}}{l_p^2}$. The log of the
degenracy is thus
\be
S_{BH}= \frac{A_{H}}{l_p^2} \ln (2j_e+1)
\label{eq}
\ee

The entropy calculation is exact, and there are no corrections to the Bekenstein-Hawking term.

In the situation one fixes the number of edges a priori, to be a number N and 
these edges are allowed to be distributed asymmetrically, i.e. the spins
$j_e$ of the edges need not be equal, the constraint is $\sum_{j_e} (j_e+1/2)= \frac{A_{H}}{l_p^2}$ the entropy is
\be
S_{BH}=\frac{A_H}{l_p^2}\left(\frac32 \ln 3 -\ln 2\right) - \frac12\ln(\frac{A_H}{l_p^2}) +..
\label{eq1}
\ee

As seen above, the $\log$ area correction appears here with the coefficient
$1/2$, and this type of correction has been obtained in other derivations
of entropy \cite{log}. However as we show below, this is not unique, and entropy
corrections will differ if the number of edges is allowed to vary.

We then generalised the case of one graph coherent state to the sum over
graphs \cite{adght} `generalised coherent state'.
Keeping the
spherical symmetry in place, the graph at the horizon can vary as per
(i) the number of edges crossing the horizon (ii) the distribution of the
edges across the horizon. These different graphs are labeled as `minimal
graphs', as one graph {\it cannot} be obtained by subdividing the edges of the other
graph. The generalised LQG Hilbert space can be written as a direct sum of Hilbert spaces
corresponding to each minimal graph.
\be
H= \oplus_{\G} H_{\G}
\ee

In a sum over graphs situation, where the number of edges is not fixed a priori,
and the $(j_e)$ are arbitrary, subject only to the constraint that
$\sum_{j_e}(j_e+1/2)=A_H/l_p^2$, the entropy is not Bekenstein-Hawking
anymore.
Two different answers are obtained, as per the two restrictions in the tracing procedure
to obtain the generalised density matrix.
\noindent
(i)The generalised density matrix is a tensor sum over the density matrices
for each Hilbert space, the entropy is given by

\be
S_{BH}= (2\frac{A_H}{l_p^2} -1)\ln 2 + \exp(-(2\frac{A_H}{l_p^2}-1)\ln 2)\ln (\frac{A_H}{l_p^2})
\label{eq2}
\ee
where, the entropy is Bekenstein-Hawking with corrections. As the area of the
horizon increases, the corrections decrease, and the leading term indeed is
a $\log$ area correction term. However, the complete correction, is decreasing
{\it exponentially in area.}\\
(ii)The second situation arises, when the entropy is obtained, from a `generalised
coherent state' which is a superposition of all the orthogonal graph Hilbert,
coherent states. This superposition allows for transition from one graph-Hilbert
space to another.
The entropy in
this case is 
\be
S_{BH}= \ln \left(\frac{1}{\sqrt 5} \left(\frac{3+\sqrt{5}}{2}\right)^{2A_H/l_p^2} - \left(\frac{3-\sqrt{5}}{2}\right)^{2 A_H/l_p^2}\right)
\label{eq3}
\ee
For large areas the leading term is indeed Bekenstein-Hawking, and the
corrections are all exponentially decreasing in area.
\be
S_{BH}= 2\frac{A_H}{l_p^2} \ln \left(\frac{3+\sqrt{5}}{2}\right) + (6.854)^{-2A_H/l_p^2}+.
\label{eq4}
\ee

\noindent
Thus, even semiclassically, the different ways of counting give different
corrections,though the leading term is universally acknowledged to be
Bekenstein-Hawking term. In particular for this derivation, the entropy
is different, as per the 'coherent state' used to describe the same 
classical space-time. If the coherent state is taken from one-graph
Hilbert space, there are no correction terms (\ref{eq}). If the coherent state
is in a superposed state as a sum over different graph coherent states,
then the correction term is exponentially decreasing in area (\ref{eq4}). 

\noindent
Thus the correction terms and the proportionality constant (which is fixed to $1/4l_p^2$ using the Immirzi parameter, and
can be done in this formalism also) in the
entropy counting remain ambiguous.
A experimental verification of the correction
term will determine the specific `generalised coherent state' the system is in. 

\section{Quantum fluctuations}
In the discussion of the above sections Bekenstein-Hawking entropy is shown
to originate from the $t\rightarrow0$ limit of the coherent state. Hence the
corrections to the `classical metric'  
have not been included in the analysis. 
What happens when one starts including corrections to the classical metric? Obviously, 
the horizon is not a impervious membrane any more.
The order $t$ 
 corrections are just proportional to $t$ multiplied
by the classical values of $h_e, P^I_e$ \cite{adg1}.  
The apparent horizon equation is no longer satisfied, and the apparent
horizon operator is not vanishing in the coherent states at this order.
The apparent horizon operator gets corrected thus:
\be
<\psi|\hat H|\psi>= t\ F(h(v_1,v_2),P(v_1,v_2)) +O(t^2)
\ee
where $F$ is a function of the classical holonomy and momenta of edges linked at both the vertices $v_1$
(outside the horizon) and $v_2$ (vertices inside the horizon). Thus, information about the phase space variables will start to emerge from behind the
horizon. 
These fluctuations can be included in the classical Hamiltonian, and the density matrix evolved in time.
Such time evolutions usually lead to a thermalisation of the density matrix, and hence we are
discovering the semi-classical origin of Hawking radiation.

\section{Discussions} The perspective described in this paper,
where entropy of horizons is attributed to a `entanglement' entropy when one
traces over the semi-classical wavefunction inside the horizon appears to be
a correct way to search for the origin of black hole entropy. Further work is in progress to obtain the coherent state in the physical Hilbert space. 

\noindent
{\bf Acknowledgements} I would like to thank the organisors of Theory Canada III
for giving me the opportunity to talk about this work.

\end{document}